\documentclass[twocolumn,amsmath,amssymb]{revtex4}
\usepackage{graphicx}
\usepackage{dcolumn}
\usepackage{bm}
\begin{document}

\title{Electrically tunable charge and spin transitions in Landau levels of 
interacting Dirac fermions in trilayer graphene} 
\author{Vadim M. Apalkov}
\affiliation{Department of Physics and Astronomy, Georgia State University,
Atlanta, Georgia 30303, USA}
\author{Tapash Chakraborty$^\ddag$}
\affiliation{Department of Physics and Astronomy,
University of Manitoba, Winnipeg, Canada R3T 2N2}

\date{\today}
\begin{abstract}
Trilayer graphene in the fractional Quantum Hall Effect regime displays a 
set of unique interaction-induced transitions that can be tuned entirely by the 
applied bias voltage. These transitions occur near the anti-crossing points 
of two Landau levels. In a large magnetic field ($> 8$ T) the electron-electron 
interactions close the anti-crossing gap, resulting in some unusual
transitions between different Landau levels. For the filling factor $\nu=\frac23$, 
these transitions are accompanied by a change of spin polarization of the ground 
state. For a small Zeeman energy, this provides an unique opportunity to control 
the spin polarization of the ground state by fine tuning the bias voltage. 
\end{abstract}
\maketitle

Dirac fermions in monolayer and bilayer graphene with their remarkable 
electronic properties have received extraordinary scrutiny in recent years 
\cite{Geim_Nobel,abergeletal}. In an external magnetic field, these systems 
exhibit unconventional quantum Hall effects \cite{Novo_05,Jiang_07} that are 
direct manifestations of their rather unusual band structures \cite{wallace,novo_06}.
As a consequence, the Landau level (LL) energies of these systems are very 
different from those of conventional two-dimensional electrons systems 
(2DESs). More specifically, in monolayer graphene the Landau level energies 
exhibit a square root dependence on the applied field \cite{old_magnetic}, 
while in bilayer graphene one finds a linear dependence \cite{falko}. 
On the other hand, interactions among Dirac fermions in the fractional 
quantum Hall effect (FQHE) \cite{stormer,FQHE_book} regime reveals several 
rather unexpected and intriguing effects in monolayer \cite{apalkov_06} and 
bilayer \cite{fqhe_bilayer} graphene. Recent experimental observation of the 
FQHE in monolayer graphene \cite{fqhe_graphene_1,fqhe_graphene_2} have indeed 
confirmed the important role electron-electron interactions play in these 
systems. Clearly, the dynamics of Dirac fermions are sensitive to the number 
of graphene layers present in the system and their stacking arrangements.
Quite expectedly, the attention has now shifted to the investigation of the 
electronic properties of Dirac fermions in trilayer graphene (TLG). 

A TLG consisting of three coupled graphene layers has a very unique electronic 
energy spectrum. Within the nearest-neighbor inter-layer coupling approximation 
the energy spectrum of TLG with Bernal stacking consists effectively of decoupled 
single-layer graphene and the bilayer graphene energy spectra. Therefore the TLG
allows us to study the energy spectra of both the massless and massive Dirac
fermions within a single system. In a strong perpendicular magnetic field, the LL
energy spectrum of TLG becomes a combination of Landau levels of single-layer 
and bilayer graphene \cite{multi_landau}. The spectrum exhibits many crossings of 
the Landau levels as a function of the magnetic field. At the crossing points the 
Landau levels are highly degenerate. The degeneracy is lifted when higher-order 
inter-layer coupling terms are taken into account, resulting in several unusual 
properties of the quantum Hall effect in trilayer graphene \cite{tri_qhe}.

\begin{figure}
\begin{center}\includegraphics[width=8.5cm]{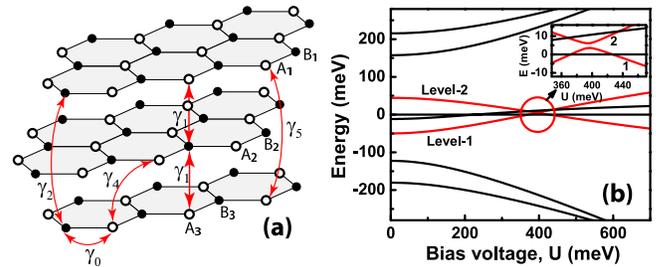}\end{center}
\vspace*{-0.5 cm}
\caption{(a) Schematic illustration of the ABA stacking of TLG. Each 
graphene layer consists of two inequivalent sites A and B. The inter-layer and 
intra-layer hopping integrals, $\gamma^{}_i$, show the couplings, which are included 
in the single-particle Hamiltonian (\ref{HABA}). (b) Landau level energy spectrum 
of TLG in a magnetic field of 15 T as a function of the bias voltage, $U$. Two red 
lines show anti-crossing for $U\approx 400$ meV. The corresponding LLs are labelled 
as level-1 and level-2, respectively. The LLs 1 and 2 belong to the set of Landau 
levels with parameter $n=0$. The inset shows the region of anti-crossing. The 
anti-crossing gap is $\approx 2.6$ meV$\approx$30 K. 
}
\label{figone}
\end{figure}

While the single-particle features of the quantum Hall effect are interesting, 
here we show that by introducing interactions among the Dirac fermions in a TLG 
in the FQHE regime, we witness several very unique properties of the TLG that 
goes far beyond the mere level crossings observed in the integer QHE. We 
found several LL repulsions and level crossings which resulted in some interesting 
spin transitions among the LLs in this system that have no analogues in the
interaction-induced spin-reversed ground states and elementary excitations 
discovered earlier in conventional electron systems \cite{spin_fqhe,tilted_expt}. 
These spin transitions in the TLG are driven by an applied perpendicular 
bias field for a fixed magnetic field, and therefore we expect these novel
transitions to be entirely tunable.

In what follows, we only consider the Bernal or ABA-stacking of our TLG. In the 
tight-binding approximation the Hamiltonian of TLG is characterized by the 
intra-layer hopping integral, $\gamma^{}_0 = 3.1$ eV, and inter-layer hopping 
integrals, $\gamma^{}_1 = 0.39$ eV, $\gamma^{}_2 = -0.028$ eV, $\gamma^{}_4 = 
0.041$ eV, and $\gamma^{}_5 = 0.05$ eV, corresponding to different types of the
inter-layer coupling, shown schematically in Fig. 1 (a) \cite{tri_qhe}. In the 
basis $(\psi^{}_{A^{}_1}- \psi^{}_{A^{}_3}, \psi^{}_{B^{}_1}-\psi^{}_{B^{}_3},
\psi^{}_{A^{}_1}+\psi^{}_{A^{}_3}, \psi^{}_{B^{}_2}, \psi^{}_{A^{}_2}, \psi^{}_{B^{}_1}
+\psi^{}_{B^{}_3})$ and in a perpendicular magnetic field, the Hamiltonian 
of a TLG for a single valley, e.g., valley $K$, takes the form 
\cite{tri_qhe,multi_landau}
\begin{widetext}
\begin{equation}
{\cal H} = \left(
\begin{array}{cccccc}
 - \gamma^{}_2/2  &  v^{}_{0} \pi^{}_{+} & -U/2 & 0 & 0 & 0 \\
 v^{}_{0} \pi^{}_{-} & -\gamma^{}_5/2 +\delta & 0 & 0 & 0 & -U/2 \\
 -U/2 & 0 &   +\gamma^{}_2/2  & 0 & - \sqrt{2} v^{}_4 \pi^{}_{+} & v^{}_{0}  
 \pi^{}_{+}  \\
 0 & 0 & 0 & 0 & v^{}_{0} \pi^{}_{-} & -\sqrt{2} v^{}_4 \pi^{}_{-} \\
 0 & 0 & -\sqrt{2}v^{}_4 \pi^{}_{-} & v^{}_{0} \pi^{}_{+} & \delta & \sqrt{2}
\gamma^{}_1 \\
 0 & -U/2 &  v^{}_{0} \pi^{}_{-} & -\sqrt{2} v^{}_4 \pi^{}_{+} & \sqrt{2} \gamma^{}_1 
& \gamma^{}_5/2 +\delta
 \end{array}
\right),
\label{HABA}
\end{equation}
where $v^{}_0 = (\sqrt{3}/2) a\gamma^{}_0/\hbar \approx 10^6 $ m/s, $v^{}_4 
= (\sqrt{3}/2) a\gamma^{}_4/\hbar $, and $\pi^{}_{\pm }=\pi^{}_x \pm \pi^{}_y$. 
Here $\vec{\pi} = \vec{p} + e\vec{A}/c$ is the generalized momentum. The parameter 
$\delta = 0.046$ eV is the difference between the on-site energies of two sublattices 
within a single graphene layer \cite{tri_qhe}. The bias voltage, $U$, is introduced 
in the Hamiltonian (\ref{HABA}) as the potential difference, i.e., the on-site energy 
difference, between layers 1 and 3. Here we assume that the potential of layer 1 
is zero, while the potential of layer 2 and 3 are $U/2$ and $U$, respectively. The 
bias voltage, $U$, is considered as the parameter of the system, which can be 
varied externally. 
\end{widetext}

The LLs of a TLG can be obtained from the Hamiltonian matrix (\ref{HABA}). The 
corresponding wave functions are parametrized by the integer $n$ and can be 
expressed through the conventional ({\it non-relativistic}) Landau level wave 
functions, $\phi^{}_{n,m}$ as
\begin{equation}
\Psi =
\left(\begin{array}{c}
 C^{}_1 \phi^{}_{n+2,m} \\
 C^{}_2 \phi ^{}_{n+1,m} \\
 C^{}_3 \phi^{}_{n+2,m} \\
 C^{}_4 \phi^{}_{n,m}  \\
 C^{}_5 \phi^{}_{n+1,m}  \\
 C^{}_6 \phi^{}_{n+1,m}
\end{array}
 \right),
\label{fABA}
\end{equation}
where $m$ is the intra-Landau level parameter, e.g., the angular momentum, and 
$C^{}_i$ are constants. Therefore, the LL wave functions of a TLG are combinations 
of $n$, $n+1$, and $n+2$ non-relativistic Landau functions. 

\begin{figure}
\begin{center}\includegraphics[width=9.0cm]{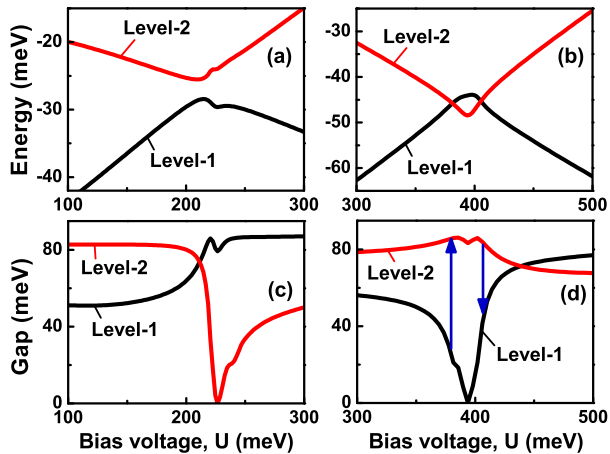}\end{center}
\vspace*{-0.5 cm}
\caption{Numerical results for finite-size $\nu=\frac13$-FQHE systems in LL 1 
and LL 2, shown as a function of the bias voltage near the anti-crossing points. 
The magnetic field is 5 Tesla [panels (a) and (c)] and 15 Tesla [panels (b) 
and (b)]. Panels (a) and (b) show the ground state energy per particle, while
panels (c) and (d) show the excitation gaps at the corresponding energy 
levels with filling factor $\nu=\frac13$. Blue arrows in panel (d) indicate
jumps of the FQHE gaps at the level crossing in panel (b). The number of electrons 
is $N=9$ and the parameter of the sphere is $2S=24$. 
}
\label{figtwo}
\end{figure}

In what follows, we consider a many-electron system partially occupying a single 
LL of the TLG. We study the properties of these systems in the FQHE regime, 
specifically for the filling factors, $\nu = \frac13, \frac23$, and $\frac25$. 
In these cases, the conventional non-relativistic system shows incompressible 
behavior with a finite energy gap \cite{stormer,FQHE_book}. The interaction 
properties of the many-electron system occupying a single Landau level are 
completely determined by the Haldane pseudopotentials $V_m^{(n)}$ \cite{haldane}, 
which are the interaction energies of two electrons with relative angular momentum, 
$m$. They are determined by making use of the LL wave functions (\ref{fABA})  
\cite{chapter}.
We numerically evaluate the FQHE state in a TLG by considering a finite-size system 
of $N$ electrons in a spherical geometry \cite{haldane} with interaction potentials 
determined by the Haldane pseudopotentials. The radius of the sphere is $\sqrt{S}
\ell^{}_0$, where $2S$ is the number of magnetic fluxes through the sphere in units 
of the flux quantum, and $\ell^{}_0=\sqrt{e\hbar/c B}$ is the magnetic length. The 
parameter $S$ also determines the number of single-particles states, $2S+1$, and for 
finite number of electrons -- the filling factor of the system. 

From the Hamiltonian (\ref{HABA}) we numerically evaluate the single-particle LL
energy spectrum. The TLG LLs are parametrized by the integer $n$ [see 
Eq.~(\ref{fABA})]. For each $n$ there are 6 LLs in a TLG. A typical LL spectrum
is shown in  Fig.~1 (b). The spectrum as a function of the bias voltage (or the magnetic 
field) shows crossing and anti-crossing of the energy levels. The anti-crossing gap, 
shown in Fig.~1 (b), is around 2.6 meV$\approx 30 $ K for a magnetic field of 15 Tesla. 
Below we show that the FQHE in TLG has non-trivial and unique 
interaction-induced properties near these anti-crossing points. We study the 
behavior of the system near the special anti-crossing point shown as inset in
Fig.~1 (b). This point corresponds to anti-crossing of the TLG LLs with $n=0$. 
We label the corresponding levels as Landau level-1 (LL-1) and Landau level-2 
(LL-2) [Fig.~1 (b)] and consider the FQHE states only in these levels. 
For each FQHE state in the corresponding LL-1 and LL-2 we have evaluated the ground state 
energy per particle and the excitation gap for the incompressible states. 

We first consider the fundamental $\nu =\frac13$-FQHE states in LLs 1 and 2, 
respectively [Fig.~2]. The system behaves very differently for weak 
and strong magnetic fields. In a weak magnetic field ($B=5$ T) the many-particle 
states show anti-crossing (Fig.~2a) similar to the single-particle levels 
[Fig.~1 (b)]. This anti-crossing is clearly visible in the dependence of the 
$\frac13$-FQHE gaps since the values of the FQHE gaps in LL-1 and LL-2 are 
interchanged when the system goes through the anti-crossing point. The system shows 
an interesting behavior exactly at the anti-crossing point. Here, due to a mixture 
of the single-particle wavefunctions of LL-1 and LL-2, the many-particle interaction 
properties are enhanced in LL-1 while suppressed in LL-2. As a result, at the 
anti-crossing point, the $\frac13$-FQHE gap in LL-1 has a maximum while the 
$\frac13$-state in LL-2 becomes compressible with a vanishing gap [Fig.~2 (c)]. 
Due to a larger cohesive energy of the incompressible state compared to the compressible 
one, the many-particle anti-crossing gap shows a small enhancement relative to the
single-particle value by $0.4 $ meV $\approx 5$ K. Experimentally, the anti-crossing 
properties of TLG in a small magnetic field can be observed by studying the FQHE in 
LL-2. In such a system, with increasing bias voltage one would observe a
transition FQHE -- no FQHE -- FQHE within a single LL, just as we predicted earlier for
bilayer graphene \cite{fqhe_bilayer}. 

Near the anti-crossing point in a large magnetic field ($B=15$ T), TLG shows several 
novel features [Fig.2 (b,d)]: The anti-crossing of the single-particle energy levels 
becomes double crossings for the many-particles states. This means that the cohesive 
energy of the many-particle state in LL-2 is larger than that in LL-1, and this 
difference overcomes the anti-crossing gap. The reason for such a behavior is the 
change in the interaction strength in LL-1 and LL-2. For a large magnetic field 
the many-particle interaction potential at the anti-crossing point becomes {\it 
stronger} in LL-2 and {\it weaker} in LL-1, which is opposite to what we see for
a weak magnetic field [Fig.~2 (a,c)]. As a result, the FQHE gap in LL-2 has a
maximum at the anti-crossing point, while the gap in LL-1 is suppressed. 
Therefore at the anti-crossing point the $\nu=\frac13$-many-particle system is 
incompressible in LL-2 and compressible in LL-1. Since the incompressible FQHE 
state has a lower binding energy than that of the compressible state, this energy 
difference is enough to close the anti-crossing gap. Therefore, in a large magnetic 
field and as a function of the bias voltage, we should expect the following behavior: 
The FQHE system, which initially for a small bias voltage, $U<400 $ meV, is in LL-1, 
occupies LL-2 at the anti-crossing point, $U = 400$ meV, leaving LL-1 empty.  
With further increase of the bias voltage, $U > 400$ meV, the system returns to 
LL-1, while LL-2 becomes empty. These transitions between different LLs at the
anti-crossing point are accompanied by jumps in the value of the FQHE gap as 
illustrated by blue lines in Fig.~2 (d). 

\begin{figure}
\begin{center}\includegraphics[width=6.0cm]{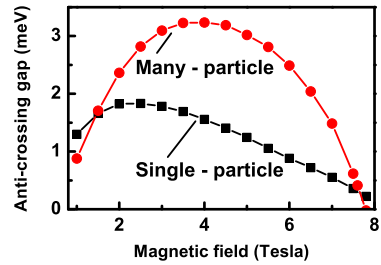}\end{center}
\vspace*{-0.5 cm}
\caption{
The anti-crossing gap corresponding to the anti-crossing of LL-1 and LL-2 
[Fig.~1 (b)], versus the magnetic field. For different magnetic fields,
anti-crossings occur for different bias voltages. The black line and the squares 
describe a single-particle system, while the red line and the circles correspond 
to the many-particle $\nu = \frac13$-FQHE system (Fig.~2). The single-particle
anti-crossing gap is closed for the many-particle system for $B\sim 8$ T. The
number of electrons in the many-electron system is $N=9$ and the parameter of the 
sphere is $2S = 24$.
}
\label{figone1}
\end{figure}

The strength of the FQHE, i.e., the magnitude of the FQHE gap, and correspondingly 
the cohesive energy of the FQHE states, is determined by the short-range properties 
of the interaction potential, i.e, the Haldane pseudopotentials at small values of 
the relative angular momentum, $m$. Therefore, at a weak magnetic field the short-range 
interaction strength at the anti-crossing point is enhanced in LL-1 (lower energy 
level), while for high magnetic fields the interaction strength at the anti-crossing 
point is increased in LL-2. This results in a weak enhancement of the many-particle 
anti-crossing gap for weak magnetic fields and strong suppression of the many-particle 
anti-crossing gap for strong magnetic fields. In Fig.~3, we present the anti-crossing 
gaps for single-particle and many-particle $\nu=\frac13$-FQHE systems for different 
magnetic fields. For small values of the magnetic fields, $1.5 {\rm T} < B< 8 {\rm
T}$, the many-particle gap clearly shows an enhancement compared to that for the 
single-particle case. For $B\approx 8$ T the many-particle anti-crossing gap closes, 
and for $B> 8$ T the anti-crossing in a single-particle system becomes a double-crossing 
in the many-particle FQHE system. 

It is known that the FQHE ground state of conventional semiconductor systems can be 
spin-polarized or spin-unpolarized. The polarization properties of the system are 
determined by the filling factor and strength of the applied magnetic field 
\cite{FQHE_book,spin_fqhe}. While in conventional systems the $\nu = \frac1m$ FQHE 
state is always fully spin-polarized, for filling factors $\nu = \frac23$ and 
$\nu = \frac25$ there is a competition between the energies of spin-polarized and 
spin-unpolarized incompressible states \cite{spin_fqhe}. With increasing strength 
of the magnetic field, the Zeeman energy of electrons favors the spin-polarized state, 
which results in possible spin-transitions in the system by varying the magnetic field. 
Those theoretical predictions subsequently received experimental confirmation 
\cite{tilted_expt,hydrostatic,opnmr,rdnmr,radiative}. The TLG system shows a different 
type of spin transitions realized at the anti-crossing points, which could also be 
probed experimentally. 

We have analyzed the spin properties of the FQHE states in TLG for filling factors 
$\nu = \frac23$ and $\nu=\frac25$. In Fig.~4 the results for $\nu=\frac23$-FQHE state 
are shown for LL-1 and LL-2 without including the Zeeman energy. The black and red 
lines in Figs.~4 (a,b) correspond to spin-polarized and spin-unpolarized systems, 
respectively. The general behavior of the system is similar to that of the $\nu = 
\frac13$-FQHE state. In a small magnetic field the system shows stronger
interactions in LL-1, while for larger magnetic fields the anti-crossing of energy levels 
becomes double crossings. For a small magnetic field, the ground state of the 
$\nu=\frac23$-FQHE system is mainly spin-polarized with only a small region of bias 
voltages, $U$, when the system becomes spin-unpolarized in LL-2. Therefore, for a weak 
magnetic field, $B\sim 5$ T, the $\nu=\frac23$-FQHE system in LL-2 should show spin 
transition into an unpolarized state within a narrow interval of $U$ at the anti-crossing 
point. A strong Zeeman energy will however suppress this spin transition. 

In a large magnetic field [Fig.~4 (b)], the $\nu=\frac23$-FQHE system shows 
interesting spin properties. While in LL-1 the $\nu=\frac23$-FQHE ground state is 
spin-unpolarized for small values of $U$, it becomes spin-polarized at a large bias 
voltage, $U > 420$ meV. In LL-2 the ground state is spin-unpolarized for all values of 
$U$. Finally, combining these two behaviors and comparing the ground state energies 
of different systems [Fig.~4 (b)], we predict the following novel spin transitions. 
If the system is initially in LL-1 for the $\nu=\frac23$-FQHE state, then with increasing 
bias voltage the system will undergo the following transitions: {\bf spin-unpolarized} 
state in LL-1 $\Leftrightarrow$ {\bf spin-polarized} state in LL-2 $\Leftrightarrow$ 
{\bf spin-unpolarized} state in LL-1 $\Leftrightarrow$ {\bf spin-polarized} state in LL-1.
What is remarkable here is that, spin polarization of the $\nu=\frac23$-FQHE system in 
TLG can be controlled by fine tuning the bias voltage -- a possibility that never existed 
in the FQHE regime of conventional systems. We have also studied the $\nu=\frac25$ filling
factor in LL-1 and LL-2. The general properties of the anti-crossing is similar to those 
of the $\nu=\frac13$-FQHE system, i.e., an enhancement of the anti-crossing gap for a 
small magnetic field and strong suppression of the anti-crossing gap for a large magnetic 
field. However, for both LL-1 and LL-2 the $\nu=\frac25$-FQHE ground state is 
spin-polarized, which excludes the possibility of spin transitions in that system. 

\begin{figure}
\begin{center}\includegraphics[width=9.0cm]{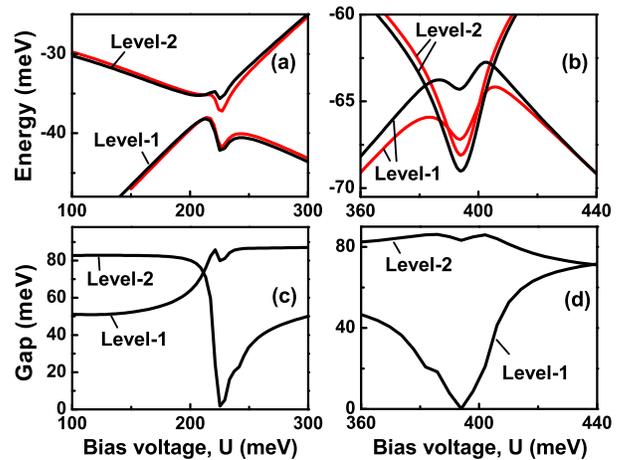}\end{center}
\vspace*{-0.5 cm}
\caption{The ground state energy per particle [panels (a) and (b)] and the excitation 
gaps [panels (c) and (d)] for $\nu=\frac23$-FQHE in LL-1 and LL-2 versus the bias 
voltage near the anti-crossing points. The excitation gaps are shown only for the 
spin-polarized systems. The magnetic field is 5 Tesla [panels (a) and (c)] and 15 Tesla 
[panels (b) and (b)]. The black and red lines in panels (a) and (b) correspond to 
spin-polarized and spin-unpolarized systems, respectively. For the spin-polarized system, 
the number of electrons is  $N=16$ and the parameter $2S$ is 24, while for the 
spin-unpolarized system $N=10$ and $2S = 12$.
}
\label{figthree}
\end{figure}

In conclusion, trilayer graphene exhibits several unique electronic properties near the 
anti-crossing points of two Landau levels. In the FQHE regime, the electron-electron 
interaction strongly renormalizes the anti-crossing gap. In a weak magnetic field 
($B<8$ T), the many-body interaction enhances the anti-crossing gap, resulting in a 
non-monotonic dependence of the excitation gaps on the bias voltage. In a large magnetic 
field ($B>8$ T), the electron-electron interaction strongly suppresses and finally 
closes the anti-crossing gap. In such large magnetic fields, the spin-polarized FQHE 
system shows nontrivial transitions as a function of the bias voltage, which are 
accompanied by jumps of the FQHE excitations gaps. In a large magnetic field ($B> 8$ T) 
the TLG displays unique spin polarizations with {\it controllable spin transitions}. 
By varying the bias voltage, the $\nu=\frac23$-FQHE system can be switched from 
spin-polarized to spin-unpolarized states. Various experimental techniques have been 
developed in the past to study spin transitions in the FQHE regime for conventional 
electron systems. These include measurements in a tilted magnetic field 
\cite{tilted_expt}, application of hydrostatic pressure \cite{hydrostatic}, optically 
pumped nuclear magnetic resonance \cite{opnmr}, resistively detected nuclear magnetic 
resonance \cite{rdnmr}, time-resolved radiative recombination \cite{radiative}, among 
others. Similar experimental studies in trilayer graphene will undoubtedly uncover a 
wealth of information about charge and spin transitions revealed by a TLG in the FQHE 
regime.

The work has been supported by the Canada Research Chairs Program of the 
Government of Canada.

\end{document}